\documentstyle[revtex]{aps}
\begin{document}
\draft
\begin{title}
Limits on Topological Defects Models of the Ultrahigh Energy Cosmic Rays
\thanks{Preprint number ADP-AT-96-6, {\it Phys. Rev. Lett.} {\bf 77}, 3708 (1996), corrected 1997.}
\end{title}
\author{R.J.~Protheroe$^1$ and Todor Stanev$^2$}
\begin{instit}
$^1$Department of Physics and Mathematical Physics, University
of Adelaide\\ Adelaide, SA 5000, Australia\\
$^2$Bartol Research Institute, University of Delaware, Newark, DE 19716, U.S.A.
\end{instit}
\begin{abstract}
  Using the propagation of ultra high energy nucleons, photons and
 electrons in the universal radiation backgrounds, we obtain limits
 on the luminosity of topological defect scenarios for the origin of
 the highest energy cosmic rays. The limits are set as function of
 the mass of the X-particles emitted by the cosmic strings or other defects,
 the cosmological evolution of the topological defects, and the strength
 of the extragalactic magnetic fields. The existing data on the cosmic
 ray spectrum and on the isotropic 100 MeV gamma-ray background
 limit significantly the parameter space in which topological defects
 can generate the flux of the highest energy cosmic rays, and rule
out models with the standard X-particle mass of $10^{16}$ GeV and higher.
\end{abstract}

\narrowtext

The cosmic ray events with the highest energies so far detected
have energies of $2 \times 10^{11}$ GeV \cite{Hay94} and 
$3 \times 10^{11}$ GeV \cite{Bir95}.
The question of the origin of these cosmic rays having energy significantly
above $10^{11}$ GeV is complicated by
propagation of such energetic particles through the universe.
The threshold for pion photoproduction on the microwave background  is 
$\sim 2 \times 10^{10}$ GeV, and at $3 \times 10^{11}$ GeV 
the energy-loss distance is about 20 Mpc.
Propagation of cosmic rays over substantially larger distances
gives rise to a cut-off in the spectrum at $\sim 10^{11}$ GeV 
as was first shown by Greisen \cite{Gre66}, and 
Zatsepin and Kuz'min \cite{Zat66}, the ``GZK cut-off''.

The standard cosmic ray acceleration mechanism, shock acceleration,
leads to a power-law energy spectrum, $dn/dE \propto E^{-\alpha}$, with
differential index $\alpha > 2$.
To reach energies of $\sim 10^{11}$ GeV one
requires the conditions present in powerful radio galaxies
\cite{RachenBiermann93}.

An alternative explanation of the highest energy
 cosmic rays is the topological defect (TD) scenario
\cite{Hill83,AharonianBhatSchramm92,BhatHillSchramm92,GillKibble94},
 where the observed cosmic rays are a result of top-down cascading,
 from 
the GUT scale energy of $\sim 10^{16}$ GeV or higher \cite{Amaldi91}, 
down to $10^{11}$ GeV and lower energies.
Generally, these models put out much of the energy in a very flat
spectrum of photons and electrons extending up to the mass of the
``X--particles'' emitted.
Approximating this spectrum by monoenergetic injection of photons of energy
$10^{15}$ GeV, Protheroe and Johnson \cite{ProtheroeJohnson96} showed
that spectra from single TD sources can not explain the
2--$3 \times 10^{11}$ GeV events.

  The main problem with topological defect models is the wide range
 of model parameters in which this scenario could, in principle, be
 applied. 
 Parameters of TD scenarios include:
 mass of the X--particle, energy spectra and final state composition
 of the decay products, and cosmological evolution of the topological
 defect injection rate \cite{Sigl95,Lee96}.
The problem of propagation is more severe than for the 
case of acceleration scenarios
because most of the energy from X--particle decay
emerges in electrons, photons and neutrinos,
with only about 3\% in nucleons.
The electrons and photons initiate electromagnetic cascades in the
extragalactic radiation fields and magnetic field, resulting in
a complicated spectrum of electrons and photons which is very
sensitive to the radiation and magnetic environment.
For example, recently the HEGRA group \cite{Karle95} have placed an
upper limit on the ratio of $\gamma$-rays to cosmic rays of $\sim 10^{-2}$
at $10^5$ GeV and, using a TD model calculation 
\cite{AharonianBhatSchramm92} which neglected the IR background
and gave a higher ratio, argued that TD models were ruled out.
However, inclusion of the IR would reduce
the $10^5$ GeV $\gamma$-ray intensity to well below the HEGRA limit.

Protheroe and Johnson \cite{ProtheroeJohnsonTaup95} considered 
one set of parameters ($M_X c^2=10^{15}$ GeV, constant injection per
co-moving volume, $B=10^{-9}$ gauss) and ruled out TD as the origin
of the 2--$3 \times 10^{11}$ GeV events.
This was mainly due to the high gamma-ray intensities
at observable energies in the electromagnetic cascade initiated by
electrons and photons in the TD spectrum above $10^{11}$ GeV.
The unification mass obtained from an analysis of LEP data \cite{Amaldi91} 
is $10^{16.0 \pm 0.3}$ GeV, and the X-particle mass cannot be far from this.  
For an X-particle mass close to the unification mass, i.e. higher than the 
$10^{15}$ GeV used in ref.~\cite{ProtheroeJohnsonTaup95}, even more energy 
would be injected into this cascade, and the gamma-ray intensities would 
violate the observational contraints even more.
Ref.~\cite{ProtheroeJohnsonTaup95} has therefore already ruled out 
TD as the origin of the 2--$3 \times 10^{11}$ GeV events.
Recently, however, Lee \cite{Lee96} and Sigl, Lee and Coppi 
\cite{SiglLeeCoppi96} have claimed that lower X-particle masses are possible,
and adopting $M_Xc^2=10^{14}$ GeV, and a lower
magnetic field, suggested the TD scenario is not ruled out.
  In this letter we consider several TD scenarios 
to put limits on the luminosity of the particle fluxes
 injected by topological defects as a function of the X--particle mass,
 the cosmological evolution of the topological defects and the
 strength of the extragalactic magnetic field, and 
consider for what range of parameters TD could explain the 
2--$3 \times 10^{11}$ GeV events.
We confirm the conclusion of Protheroe and Johnson 
\cite{ProtheroeJohnsonTaup95} that for X--particle masses
of $10^{15}$ GeV or higher
TD can not explain the 2--$3 \times 10^{11}$ GeV events and
severely limit models with lower $M_X$.

We use the same injection spectra and TD evolution as in 
ref.~\cite{SiglLeeCoppi96}.
This is approximately an $E^{-1.5}$ spectrum extending up to
$\sim M_Xc^2/2$ contining $\sim 3$\% nucleons and 97\% pions.
In the matter dominated era of the universe, and assuming $q_0=0.5$,
the injection rate per co-moving volume is
$Q(t) = Q_0 (t/t_0)^{-2+p}$
where $p=1$ for ordinary cosmic strings and monopole-antimonopole annihilation,
$p=0$ for superconducting cosmic strings, and $p=2$ for models with
constant injection.
 
We inject this spectrum at various distances and carry out a
Monte-Carlo matrix propagation calculation as described in 
ref. \cite{ProtheroeJohnson96}.
 The following  processes are included:
 $\gamma\gamma \rightarrow e^+e^-$ on the microwave, radio and
 IR/optical background, IC scattering on the same backgrounds,
 triplet pair production and double pair production 
 on the microwave background, 
 synchrotron radiation in the extragalactic magnetic field, and redshifting
 due to expansion of the universe.
 Nucleons undergo pion photoproduction
 interactions and protons undergo Bethe-Heitler
pair production in the same environment, and
 neutron production and decay are taken into account.

  For the radio background we use the spectrum of Clark et al. \cite{Cla70}.
 Other estimates of 
 the radio background based on data on radio galaxies and ordinary galaxies
 give a radio background extending to significantly lower
 frequencies \cite{Ber69,ProtheroeBiermann96}, and we shall discuss
 the effect of using different radio 
spectra elsewhere \cite{ProtheroeStanev96}.
 Magnetic field values we use are 10$^{-15}, 10^{-12}, 10^{-11}, \dots
10^{-8}$ G.
 The values at the high end of this range may be appropriate if
 topological defects are seeds
 for the formation of galaxies and larger structures in the universe  
 \cite{Brandenburger96} where fields are generally higher than average.
For the infrared background we adopt a spectrum \cite{ProtheroeStanev93}
which is based on the model of Stecker et al. \cite{Stecker93} but
constrained at low frequencies by upper limits derived by us from the
error bars on the microwave background measured by the FIRAS experiment 
on COBE \cite{Mather94}.
At $3 \times 10^{-3}$ eV, where the microwave background is
decreasing rapidly with energy, our IR spectrum is a factor of 5 lower than 
that used by Lee \cite{Lee96}.

For a uniform distribution of topological defects we obtain the 
total intensity by integrating over redshift the
results obtained for propagation over fixed distances, taking account of
topological defect evolution and cosmological expansion
assuming $H_0=75$ km s$^{-1}$ Mpc$^{-1}$ and $q_0=0.5$.
The result for  $M_Xc^2 = 10^{14.1}$ GeV, a magnetic field of $10^{-9}$ gauss,
and $p=2$, is shown in Fig.~\ref{fig1} where 
we have normalized the spectrum of ``observable particles''
(nucleons, photons, electrons) to the $3 \times 10^{11}$ GeV point
(cosmic ray data are taken from \cite{Stanev92}, and
the highest point is from \cite{Bir95}).
Lee \cite{Lee96} has published a spectrum for similar input parameters
and it is in acceptable agreement with the present work
except for MeV--PeV $\gamma$-rays where our result is about a factor
of 10 lower.
We suspect this is because of our lower IR field, and this appears
to be confirmed by results presented by Lee which show the
$\gamma$-ray intensity to be significantly lower if the IR field
is neglected.
With our lower IR field, and consequent lower $\gamma$-ray
intensity, we are less likely
to rule out topological defect models due to excess $\gamma$-ray production.
For normalization to the $3 \times 10^{11}$ GeV data, 
the injection rate of energy in X--particles
would be $\sim 6 \times 10^{-35}$ erg cm$^{-1}$ s$^{-1}$.
Notice that above $10^{11}$ GeV photons dominate the spectra
of observable particles, and that over some ranges of energy
electrons dominate the electromagnetic component.
Also note that the predicted $\gamma$-ray flux at GeV energies is
comparable to the observed background, and as pointed out by Lee \cite{Lee96},
the extragalactic $\gamma$-ray background at these energies will place
a strong constraint on the topological defect models.
Figure~\ref{fig2} shows the energy injection
rate at the present epoch,
such that the  intensity of observable particles is normalized to 
the $3 \times 10^{11}$ GeV point, as a function of $M_X$ for 
various extragalactic magnetic fields and evolution models.

Synchrotron radiation is very important in determining the $\gamma$-ray
spectrum at MeV--PeV energies which can vary by orders of magnitude
depending on the magnetic field.
Limits on the injection rate obtained
from comparing the predicted 0.1--10 GeV intensities with 
SAS-II \cite{ThompsonFichtel} and preliminary EGRET \cite{Fichtel96} data are
only lower than the injection rate obtained from normalization 
at $3 \times 10^{11}$ GeV (and thus rule out a TD origin for the 
2--$3 \times 10^{11}$ GeV events) for the highest magnetic fields.
Where this limit is lower than the injection rate obtained from normalizing 
the intensity of observable particles to the $3 \times 10^{11}$ GeV point,
these limits have been added to Figure~\ref{fig2} 
for the three evolution models.
We see that the $\gamma$-ray data provide the strongest constraint for 
models with  high $M_X$, high $B$ and weak evolution.
No models with $M_Xc^2 < 10^{14.4}$ GeV are excluded by the constraints
imposed so far, so if we used only these two constraints we would
agree with Sigl et al. \cite{SiglLeeCoppi96} that TD scenarios 
are not ruled out.

A further constraint, not considered by Sigl et al. \cite{SiglLeeCoppi96},
 comes from the intensity of potentially 
observable particles above $3 \times 10^{11}$ GeV.
This constraint has already been used by Protheroe and Johnson
\cite{ProtheroeJohnsonTaup95} to rule out the model with $M_X=10^{15}$ GeV,
$p=2$, and $B=10^{-9}$ gauss.
Here we use the fact that 1 event was observed by the Fly's Eye between
$10^{11.45}$ GeV and $10^{11.55}$ GeV, together with the published
intensity at $10^{11.5}$ GeV, to obtain the exposure factor 
of the Fly's Eye experiment at this energy.
Assuming the exposure factor has the same value also at higher energies
(a reasonable assumption as at these energies optical transmission
will limit the distance to observable air showers rather than the 
inverse-square law), we can estimate the number of events which
should have been observed above $3 \times 10^{11}$ GeV.
Given that no events have been seen above this energy, we set a 90\%
upper limit of 2.3 events which, when compared with the expected number
of events, sets a new upper limit to the rate of injection of energy
in X-particles.
This limit is approximately independent of topological defect 
evolution and is plotted 
in Figure~\ref{fig3} against $M_X$ for various magnetic fields.
In {\em all cases} this limit is lower than the injection rate
required to explain the 2--$3 \times 10^{11}$ GeV events, and so it would 
appear that, subject to $\gamma$-rays and electrons above this energy
being detectable by the Fly's Eye as discussed below, topological defect
models are ruled out as the explanation of the 
2--$3 \times 10^{11}$ GeV events.
Comparing Figs.~2 and 3, and extrapolating to
$10^{16}$ GeV, it is obvious that TD models 
with standard $M_X$ are also ruled out.

The limits on the injection rate of energy in X--particles
from the number of ``observable particles'' above $3 \times 10^{11}$ GeV 
may actually be weaker than given in Figure~\ref{fig3} because these particles
are dominated by photons and electrons which might be undetectable.
Energetic $\gamma$-rays entering the atmosphere will be subject to the
LPM effect \cite{LPM} (the suppression of electromagnetic cross-sections
 at high energy) which becomes very important. 
The radiation length changes as $(E/E_{\rm LPM})^{1/2}$, where 
$E_{\rm LPM} = 6.15 \times 10^4 \ell_{\rm rad}$ GeV, and 
$\ell_{\rm rad}$ is the standard Bethe-Heitler radiation length
in cm~\cite{StanevLPM82}.
We find that average shower maximum will be reached 
 below sea level for energies $5 \times 10^{11}$ GeV, $8 \times 10^{11}$ GeV, 
and $1.3 \times 10^{12}$ GeV for gamma--rays entering the atmosphere
 at $\cos{\theta}$ = 1, 0.75, and 0.5 respectively. 
Such showers would
 be very difficult to reconstruct by experiments such as Fly's Eye
and at best would be assigned a lower energy.
In this case, we should treat electrons and $\gamma$-rays as unobservable,
and normalize the {\em nucleon intensity} to the $3 \times 10^{11}$ GeV data.
This has the effect of increasing the predicted $\gamma$-ray intensities,
and the new upper limits to the rate of injection of energy in X--particles
would be as given in Figure~\ref{fig4}.
We now see that normalizing to the $3 \times 10^{11}$ GeV data violates the
$\gamma$-ray data for all models with $p=2$, models with $p=1$ and 
$M_X > 10^{13.1}$--$10^{13.7}$ GeV depending on $B$, and
models with $p=0$ and 
$M_X > 10^{13.9}$--$10^{14.9}$ GeV depending on $B$.
Thus, models with standard $M_X$ would also be ruled out as the 
explanation of the 2--$3 \times 10^{11}$ GeV events.

Before entering the Earth's atmosphere  
$\gamma$-rays and electrons are likely to interact on the geomagnetic
 field (see Erber \cite{Erber68} for a review of the
 theoretical and experimental understanding of the interactions).
 In such a case the $\gamma$-rays propagating perpendicular to the geomagnetic
 field lines would cascade in the geomagnetic field, i.e. pair production
 followed by synchrotron radiation. 
The cascade process would degrade the $\gamma$-ray energies to some extent
(depending on pitch angle), and the atmospheric cascade would then
 be generated by a bunch of $\gamma$-rays of lower energy. 
Aharonian et al. \cite{AharonianKanevskySahakian91} have
considered this possibility and conclude that this bunch would
appear as one air shower made up of the superposition of many sub-showers of 
lower energy where the LPM effect is negligible, the air shower 
having the energy of the initial $\gamma$-ray outside the geomagnetic field.
If this is the case, then $\gamma$-rays above $3 \times 10^{11}$ GeV
would be observable by Fly's Eye, etc., and the upper limits presented
in Figure~\ref{fig3} would stand, ruling out a TD origin for the
2--$3 \times 10^{11}$ GeV events.
There is however, some uncertainty as to whether pair production will take 
place in the geomagnetic field.
This depends on whether the geomagnetic field spatial
dimension is larger than the formation length of the electron
pair, i.e. the length required  to achieve a separation between the
two electrons which is greater than the classical radius of the electron. 
This question of whether or not pair production in the geomagnetic field 
takes place needs further investigation.
In any case, we find TD models
for the 2--$3 \times 10^{11}$ GeV events are ruled out for 
standard X-particle masses of $10^{16}$ GeV or higher, and 
our results severely constrain models with lower $M_X$.

The research of RJP is supported by the Australian Research Council, and
the research of TS is supported in part by the US DOE under contract  
DE-FG02-91ER40626.

\newpage
 ~\\
\vspace{7.3cm}
\figure{
\includegraphics{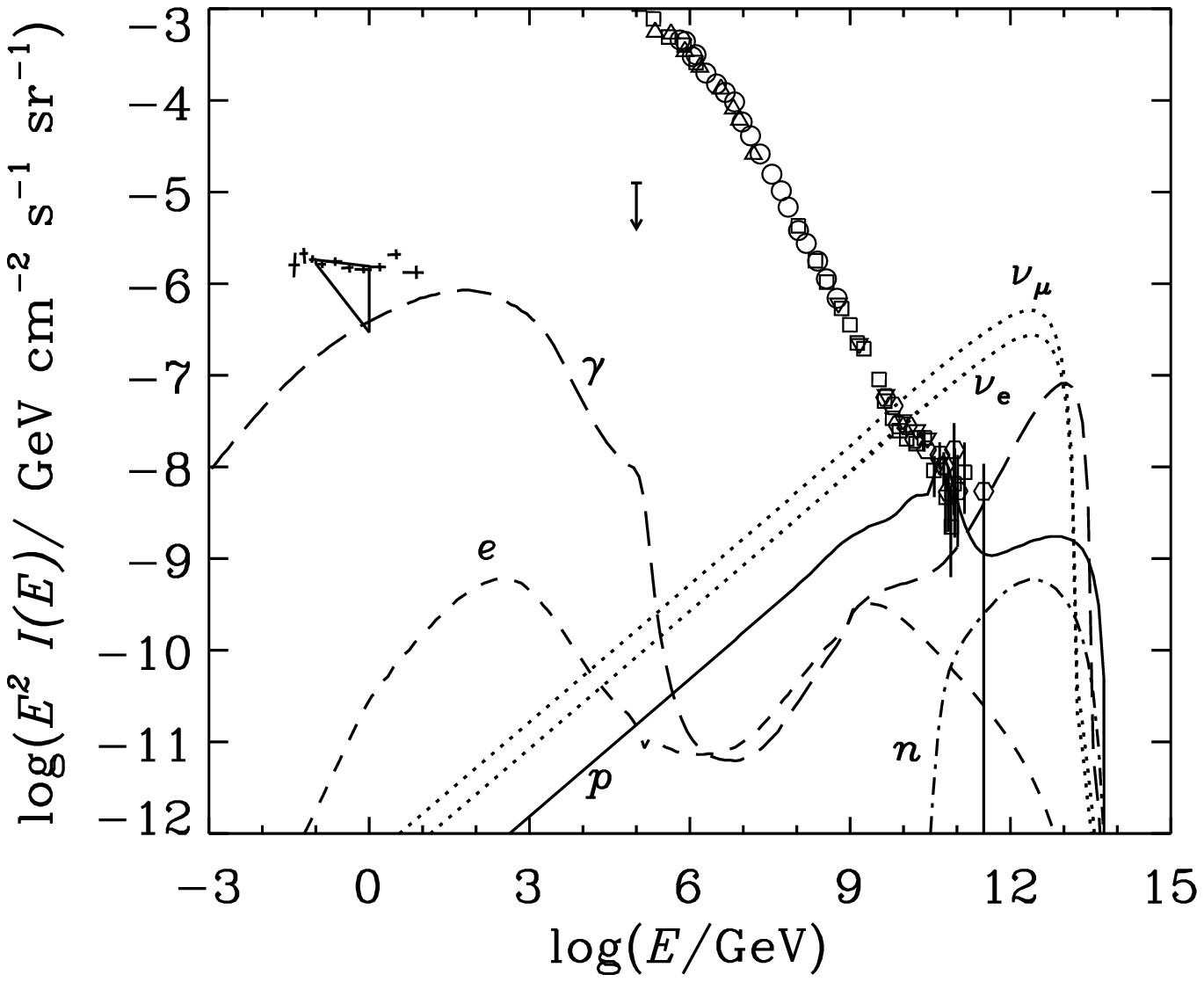}
Spectra at Earth for the topological defect model discussed in the 
text.  SAS-2 and EGRET 
$\gamma$-ray data are shown at GeV energies, and HEGRA data at 100 TeV.
\label{fig1}}

\vspace{7.3cm}
\figure{
\includegraphics{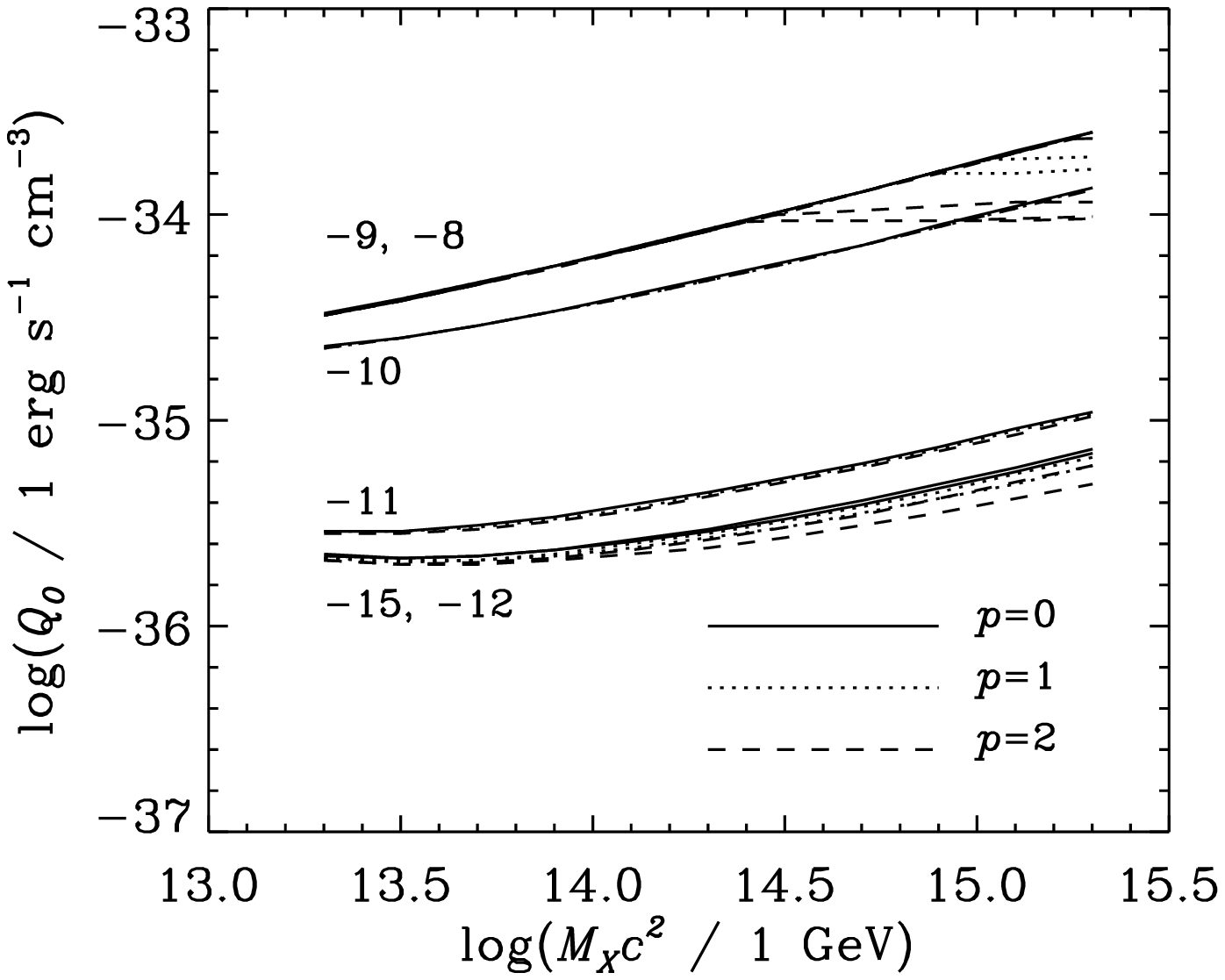}
Maximum rate of injection of energy in X--particles as a function
of $M_X$ for various magnetic fields and evolution models
based on normalization of predicted intensity of ``observable particles''
to the $3 \times 10^{11}$ GeV point, or using the $\gamma$-ray data as 
upper limits (the lower of the two is plotted).
Numbers attached to curves give log($B$/~1 G).
\label{fig2}}

\newpage
 ~\\
\vspace{7.3cm}
\figure{
\includegraphics{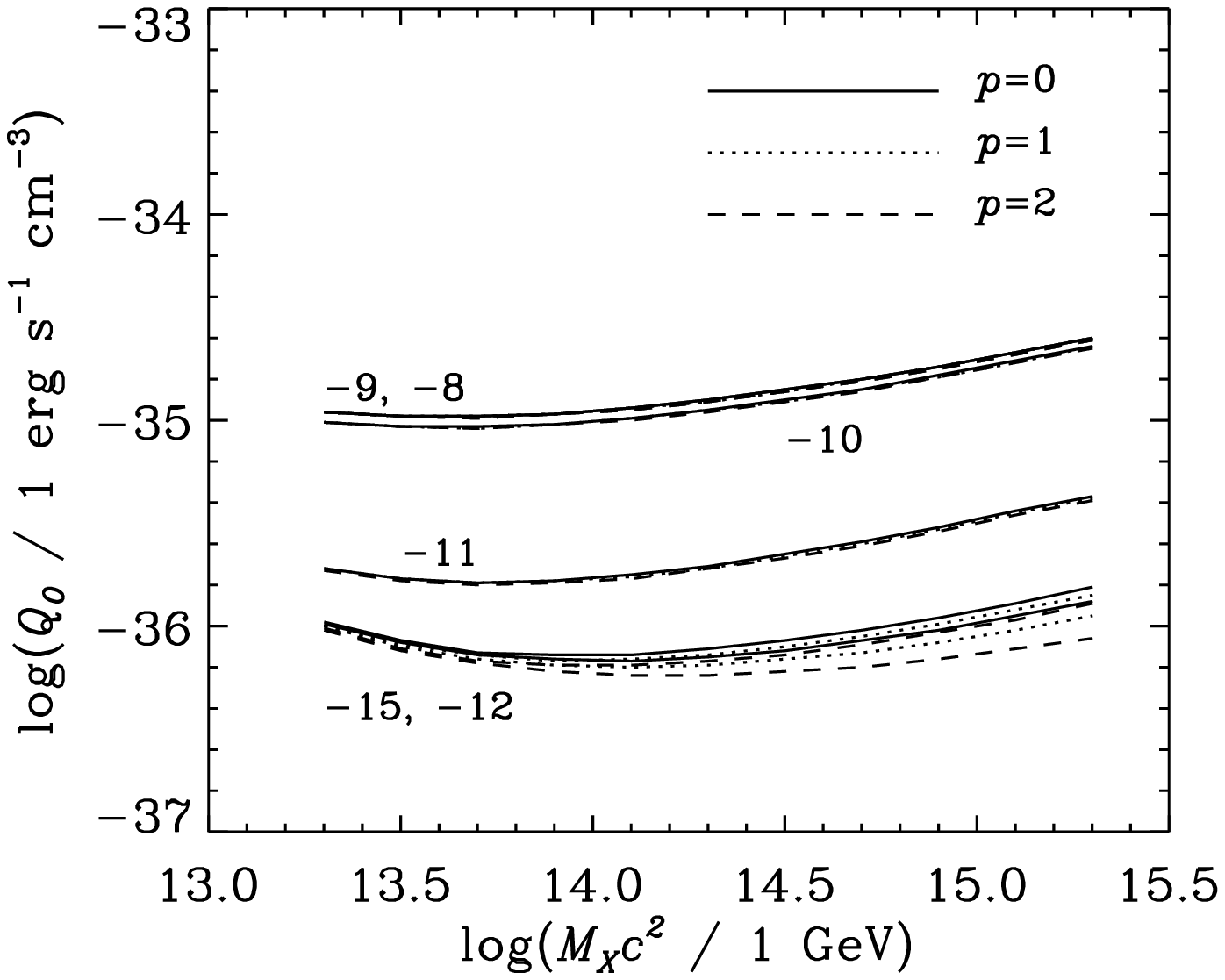}
Maximum rate of injection of energy in X--particles as a function
of $M_X$ for various magnetic fields and evolution models
based on the non-observation of cosmic rays above $3 \times 10^{11}$ GeV.
\label{fig3}}

\vspace{7.3cm}
\figure{
\includegraphics{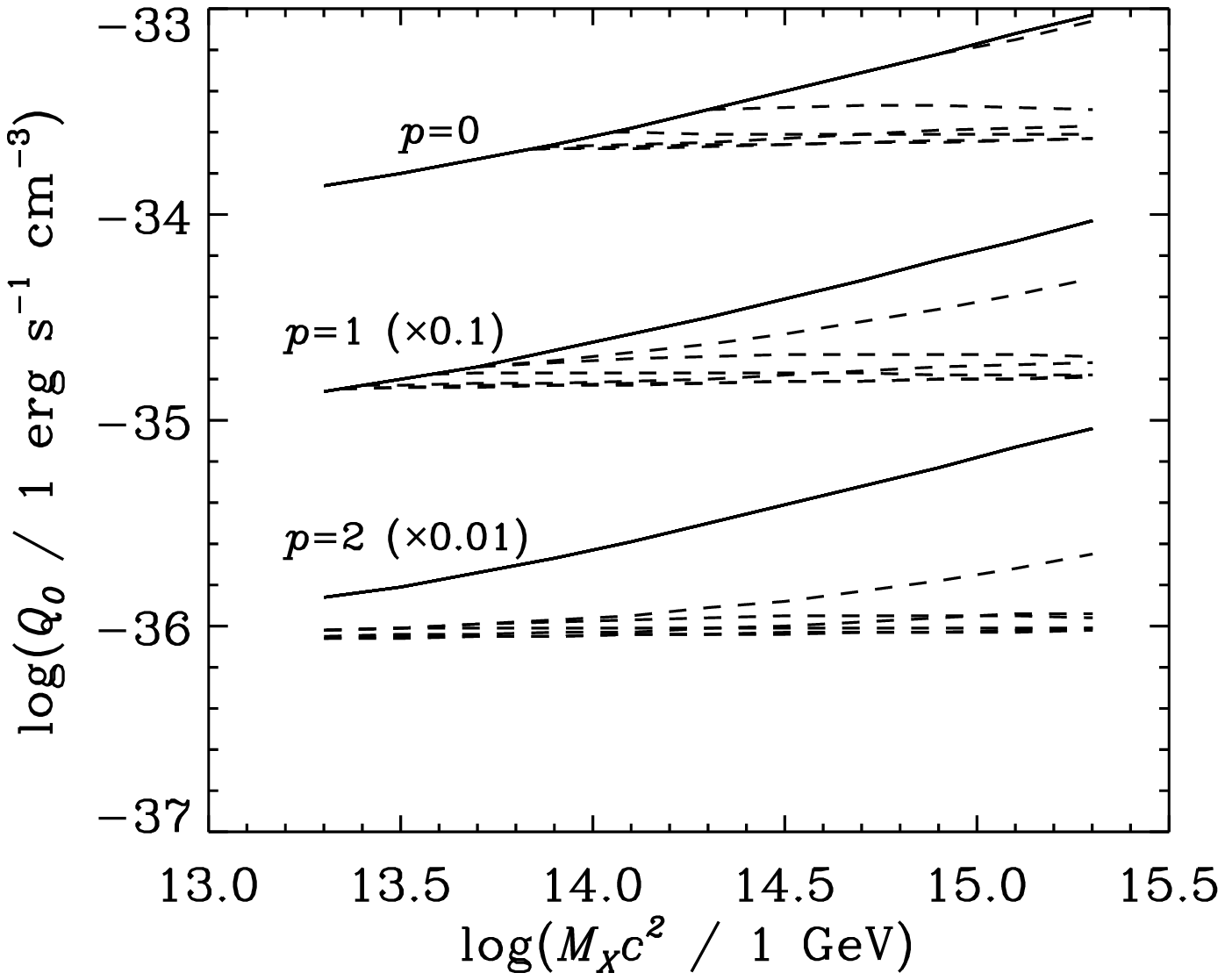}
Maximum rate of injection of energy in X--particles as a function
of $M_X$ for various magnetic fields and evolution models
based on normalization of predicted intensity of nucleons (solid curves),
and using the $\gamma$-ray data as upper limits (dashed curves) for 
$B=10^{-15}$ G (highest curves) to $B=10^{-8}$ G (lowest curves).
\label{fig4}}

\end{document}